# First-Principles Thermodynamic Analysis of Ternary Chalcogenide Phase Change Materials

Felix Adams [a], Ichiro Takeuchi [a], Carlos A. Ríos Ocampo [a,b], Yifei Mo [a,c *]

[a] Department of Materials Science and Engineering, University of Maryland, College Park, MD 20742

[b] Institute for Research in Electronics and Applied Physics, University of Maryland, College Park, MD 20742

[c] Maryland Energy Innovation Institute, University of Maryland, College Park, MD 20742

* Email: yfmo@umd.edu

**Abstract.** Chalcogenide phase-change materials (PCMs) are important for nonvolatile memory and reconfigurable photonic technologies. The GeTe–$Sb_2Te_3$ system, commonly referred to as GST, is the best-known PCM family, but new PCMs are needed to broaden the accessible property space while retaining fast and reversible switching. Here, we propose a thermodynamic framework, motivated by Ostwald's rule, for understanding and identifying PCMs, since direct modeling of phase-transition dynamics is computationally expensive. Using first-principles calculations, we systematically evaluate the energetics of ternary chalcogenide mixtures and their polymorphs along binary-binary tie lines. By comparing ground-state and metastable structures, we assess phase stability, miscibility, and the likelihood of GST-like polymorph-mediated crystallization pathways across a broad range of ternary chalcogenide mixtures. The calculations reproduce known behavior in GST and related systems and identify several promising candidate mixtures with similar features. These results provide insight into why some PCM systems are more favorable than others and establish a thermodynamic framework for future PCM discovery.

# 1 Introduction

Phase-change materials (PCMs) based on chalcogenide semiconductors can reversibly switch between amorphous and crystalline states while exhibiting large and nonvolatile changes in electrical resistance and optical properties[1,2]. These materials are used in nonvolatile memory and reconfigurable photonic technologies[1–3]. For many applications, rapid amorphous-to-crystalline phase-transition dynamics are essential because they limit device speed. Emerging applications place additional demands on PCM materials. For example, high-temperature embedded computing applications may require crystallization temperatures higher than those of current PCMs to avoid spontaneous crystallization and improve device lifetime[1]. Novel photonic applications seek PCMs with switching behavior over broader wavelength ranges[4]. These needs motivate the discovery and design of new PCM systems that retain fast switching while expanding the accessible property space.

The Ge-Sb-Te ternary system, formed by the mixture along the tie line between GeTe and $Sb_2Te_3$, commonly referred to as GST, is a well-known PCM system because of its fast switching dynamics and favorable electrical and optical contrast[1,2]. Prior studies of GST have identified a metastable rock-salt (RS) phase along its crystallization pathway[5]. GST crystallizes through a two-step process in which the amorphous phase first transforms into a metastable RS structure and then into the ground-state hexagonal phase[5]. The RS phase also contains a high vacancy concentration on the cation sublattice, giving it substantial tolerance to structural and compositional disorder while potentially facilitating the amorphous-to-crystalline transition[6,7]. The existence of this intermediate RS phase has been associated with the rapid crystallization behavior of GST. This behavior can be understood through Ostwald's rule, which states that metastable polymorphs may form before the final equilibrium structure because they provide kinetically more accessible transformation pathways[8,9].

Atomistic simulations based on *ab initio* molecular dynamics have provided insight into amorphous-to-crystalline transformation mechanisms in PCMs[2,10]. However, applying such methods systematically across a broad chemical space is computationally prohibitive, often requiring tens of thousands of CPU-hours to simulate a single composition. This challenge motivates the search for simple descriptors that can identify materials systems with promising phase-transition behavior. The GST example suggests that the presence of a low-energy metastable polymorph may provide a useful thermodynamic descriptor for PCM discovery. Specifically, if a PCM contains a metastable polymorph whose energy lies only slightly above that of the stable crystalline phase, that polymorph may serve as a kinetically favorable intermediate along the crystallization pathway. Density functional theory (DFT) calculations predict that the RS polymorph of GST lies less than 3 meV/atom above the stable hexagonal phase[11], supporting this thermodynamic interpretation. This observation motivates a simple hypothesis: materials containing low-energy metastable polymorphs may be more likely to exhibit GST-like polymorph-mediated crystallization pathways. Consistent with this hypothesis, a prior experimental study of the GeTe–$Bi_2Te_3$ system also reported the presence of an RS polymorph and fast phase-transition behavior[12]. This thermodynamic framework may therefore be used to identify new PCM systems, particularly because many chalcogenide compounds exhibit polymorphism.

PCM discovery remains largely focused on chemical modifications of existing systems, such as doping known compositions or substituting one end-member compound with chemically related analogs[13,14]. The thermodynamics of mixtures across the broader chalcogenide PCM space remain poorly understood. In particular, there has been no systematic assessment of mixtures along binary-binary tie lines with respect to phase stability, miscibility, and metastable polymorphs. This gap limits mechanistic understanding and rational materials discovery.

In this work, we evaluate candidate PCM systems within a thermodynamic framework motivated by Ostwald's rule interpretation of GST crystallization. Using first-principles calculations, we systematically evaluate the energetics of ternary compositions, including ground-state and metastable polymorphs, along the tie lines of binary-binary mixtures in chalcogenide semiconductors. We compare our results with available experimental literature to validate the proposed framework and rationalize why some chalcogenide mixtures are more promising PCM candidates than others. By comparing the energies of experimentally known and candidate polymorphs, we identify systems with low-energy metastable structures, such as RS-like phases in GST, that could enable fast phase transitions.

## 2 Results

Candidate ternary PCMs are derived from mixtures of binary chalcogenides. We first identify promising pairs of binary compositions to mix for our proposed framework, and then generate the corresponding derivative ternary structures over a range of compositions along the selected tie lines (Section 2.1). GST is then examined as a representative system in Section 2.2 to validate our approach and the thermodynamic framework motivated by Ostwald's rule. Sections 2.3–2.5 present the results for the remaining candidate systems.

### 2.1 Generation of Candidate Materials

| | GeTe | SnTe | $Sb_2Te_3$ | $Bi_2Te_3$ | $SiTe_2$ | $GeSe_2$ | GeSe | SnSe | $SnSe_2$ | $Sb_2Se_3$ | $Bi_2Se_3$ | $SiSe_2$ |
|---|---|---|---|---|---|---|---|---|---|---|---|---|
| GeTe | | ch | h | h | h | o | co | | | | | |
| SnTe | ch | | h | h | h | | | c | h | | | |
| $Sb_2Te_3$ | h | h | | h | | | | | | | | |
| $Bi_2Te_3$ | h | h | h | | h | | | | | | h | |
| $SiTe_2$ | h | h | | h | | | | | | | | |
| $GeSe_2$ | o | | | | | | | ot | | o | ot | mo |
| GeSe | co | | | | | | | co | | o | o | o |
| SnSe | | c | | | | ot | co | | | o | ot | o |
| $SnSe_2$ | | h | | | | | | | | | h | |
| $Sb_2Se_3$ | | | | | | o | o | o | | | o | o |
| $Bi_2Se_3$ | | | | h | | ot | o | ot | h | o | | o |
| $SiSe_2$ | | | | | | mo | o | o | | o | o | |

**Table 1. Candidate binary mixtures and their matching crystal families.** Crystal families: c=cubic, h=hexagonal, m=monoclinic, o=orthorhombic, t=tetragonal. For mixtures with multiple matching polymorphs, all matching crystal families are shown. Additional selected mixtures are included in the Supplemental Information.

We first identify prospective mixtures of binary chalcogenides in the Si, Ge, Sn, Sb, or Bi with Se or Te composition space, which includes many known PCM systems. For each binary composition, we consider only experimentally known phases or the lowest DFT-energy structures from the Materials Project (MP), and screen them according to the following criteria. Because we focus on ternary mixtures similar to GST, the two binary compounds must have one element in common. To select mixtures that are more likely to form single-phase materials like GST, at least one polymorph from each binary must belong to the same crystal family. The two binary compounds must be connected by a tie line in the MP phase diagram based on DFT energies, since otherwise phase segregation would be expected in the mixture. This procedure identifies 29 candidate mixtures (summarized in Table 1), including the well-known GeTe–$Sb_2Te_3$ (GST) system. Selected additional mixtures are shown in the Supplemental Information (SI).

In general, most tellurides have a polymorph in the hexagonal crystal family, with close-packed layered structures with octahedral cation coordination (see Figure 1c–e for the relevant GeTe and $Sb_2Te_3$ structures). In contrast, selenides generally have more structurally complex orthorhombic polymorphs with a wider range of coordination environments. For example, orthorhombic $Sb_2Se_3$ (MP ID mp-2160) has an Sb site with square pyramidal coordination, and orthorhombic $Bi_2Se_3$ (MP ID mp-23164) has Bi in square-face capped trigonal prism coordination.

Next, for each candidate mixture, ternary structures were generated over a range of compositions by interpolating between the parent binary compositions along the tie line (Figure 1a). The ternary structures were constructed from the parent binary phases using defect reactions involving cation or anion vacancies (see Methods), consistent with the high concentration of cation-site vacancies observed in the RS phase of GST over a wide composition range.

## 2.2 GST Polymorph Structures and Energies

Among the 29 candidate mixtures identified (Table 1), this procedure correctly recovers the well-known GST system, i.e., the GeTe–$Sb_2Te_3$ mixture (Figure 1a). Figure 1b shows the energies of individual $Ge_2Sb_2Te_5$ structures generated from each of the GeTe or $Sb_2Te_3$ polymorphs. Figure 1f and g show the lowest-energy structures of $Ge_2Sb_2Te_5$ derived from the RS (Figure 1c) and hexagonal (Figure 1d) parent binary structures, respectively. The lowest-energy GST structures generated by our algorithm exhibit the same vacancy ordering and similar Ge and Sb ordering as those reported by Da Silva et al.[11] , validating our structure-generation approach.

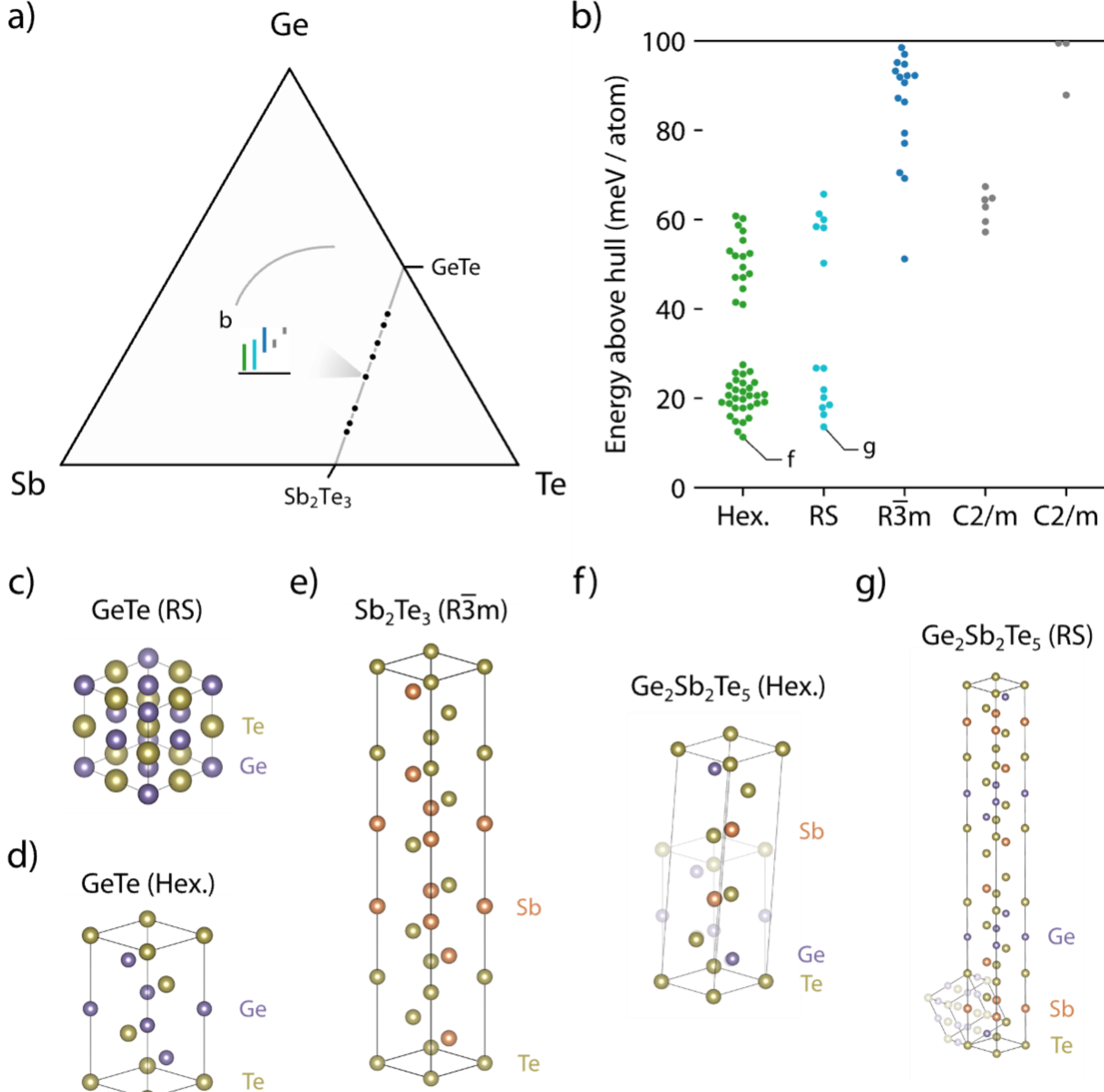

**Figure 1. Structures and energies of $Ge_2Sb_2Te_5$.** a) The Ge-Sb-Te ternary composition space, showing the GeTe–$Sb_2Te_3$ tie line and select mixing compositions. b) The energies of structures in different phases of $Ge_2Sb_2Te_5$. c) Rock salt GeTe structure. d) Hexagonal GeTe structure. e) Hexagonal $Sb_2Te_3$ structure. f) The lowest-energy generated $Ge_2Sb_2Te_5$ structure in the hexagonal phase. g) The lowest-energy generated $Ge_2Sb_2Te_5$ structure in the rock salt phase. The translucent structures overlaid on f and g are the parent structures d and c.

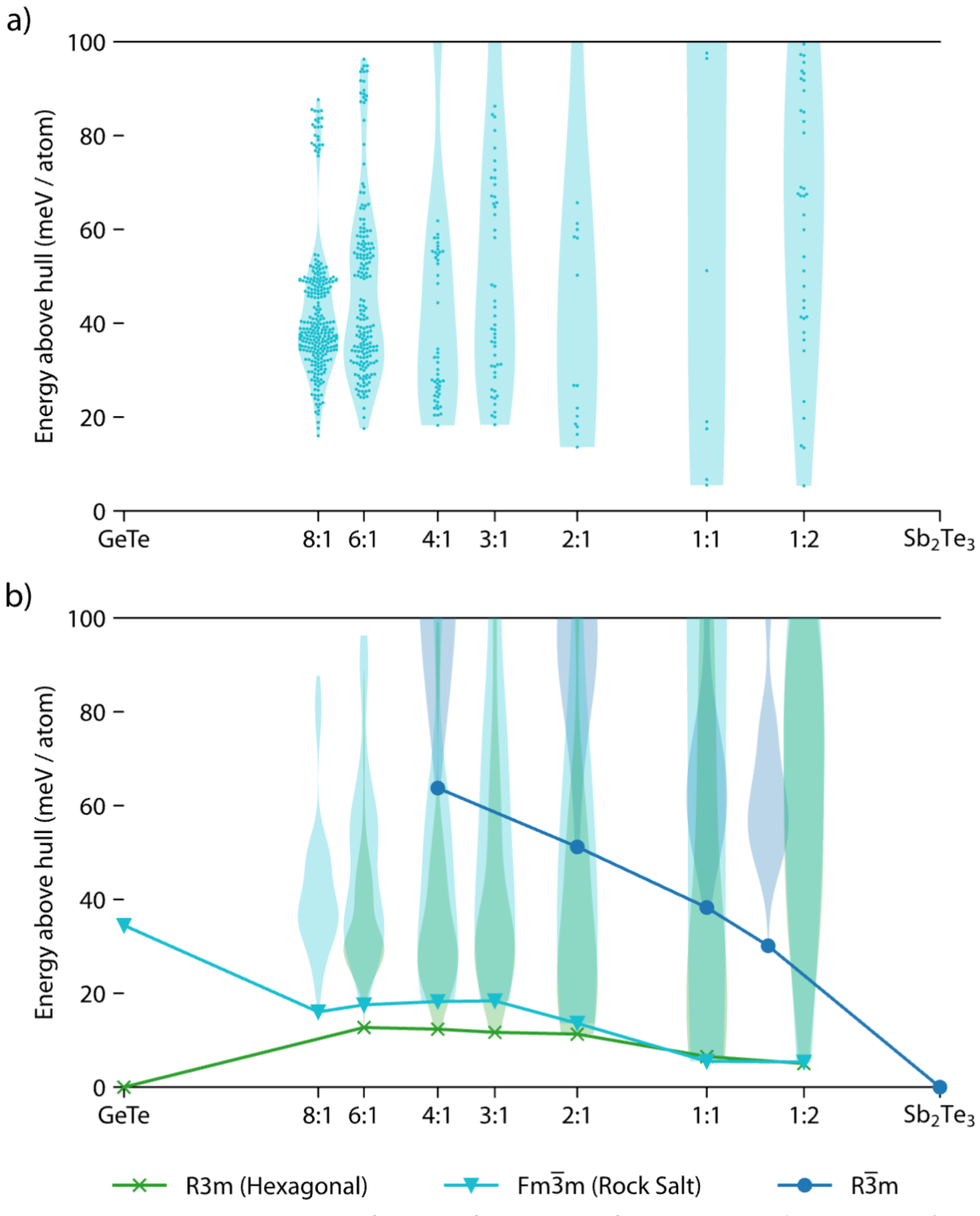

**Figure 2. DFT-calculated energy as a function of composition for GeTe–$Sb_2Te_3$.** a) The energies of ternary structures in the rock salt phase along the GeTe–$Sb_2Te_3$ tie line. Each vertical group corresponds to the composition below it, with some horizontal jitter added to prevent overlapping. The violin plots show the kernel density estimate of the energy distribution. b) The energy distributions (violin plot) and minimum energies (marker) as a function of composition for the rock salt, hexagonal, and $R\bar{3}m$ phases of GST. Energies above 100 meV/atom are not shown for clarity.

Figure 2a shows the DFT-calculated energy as a function of composition for the derived ternary structures in the RS phase of the GST system generated by our algorithm. Figure 2b compares the energy distribution and minimum energy of the RS, hexagonal, and $R\bar{3}m$ phases across the GST tie line. The lowest-energy structures of the hexagonal-phase GeTe–$Sb_2Te_3$ ternary mixtures have energies below 18 meV/atom, which is below the 25 meV/atom cutoff for ternary-mixture stability in this study. Our calculations correctly predict that mixtures of GeTe and $Sb_2Te_3$ form single-phase ternary compounds, agreeing with the experimental observation of single-phase ternary compounds, such as $Ge_2Sb_2Te_5$,[15] $Ge_1Sb_2Te_4$,[16] $Ge_1Sb_4Te_7$.[17]

The hexagonal phase is identified as the ground state, and the RS phase is identified as slightly metastable with an energy difference of $< 10$ meV/atom, in agreement with the experimental observations[5] and computational studies[11]. By contrast, the ternary mixture derived from the $R\bar{3}m$ $Sb_2Te_3$ phase is far above the 25 meV/atom cutoff. These results are consistent with experiments and support Ostwald's rule explanation for the crystallization pathway of GST, in which the metastable RS phase crystallizes first followed by the ground-state hexagonal phase[5,8,9]. The successful reproduction of the known GST polymorph energetics supports the proposed thermodynamic framework and provides a basis for its application to the discovery of new candidate PCM compositions in the systems examined below.

## 2.3 Telluride Mixtures

Here we present our calculation results for the remaining telluride candidate mixtures. Figure 3 shows the energy of different polymorphs as a function of composition for the multi-cation telluride systems. The GST system is shown in the first row, second column, as a simplified version of Figure 2b that includes only the minimum energy of each polymorph at each composition.

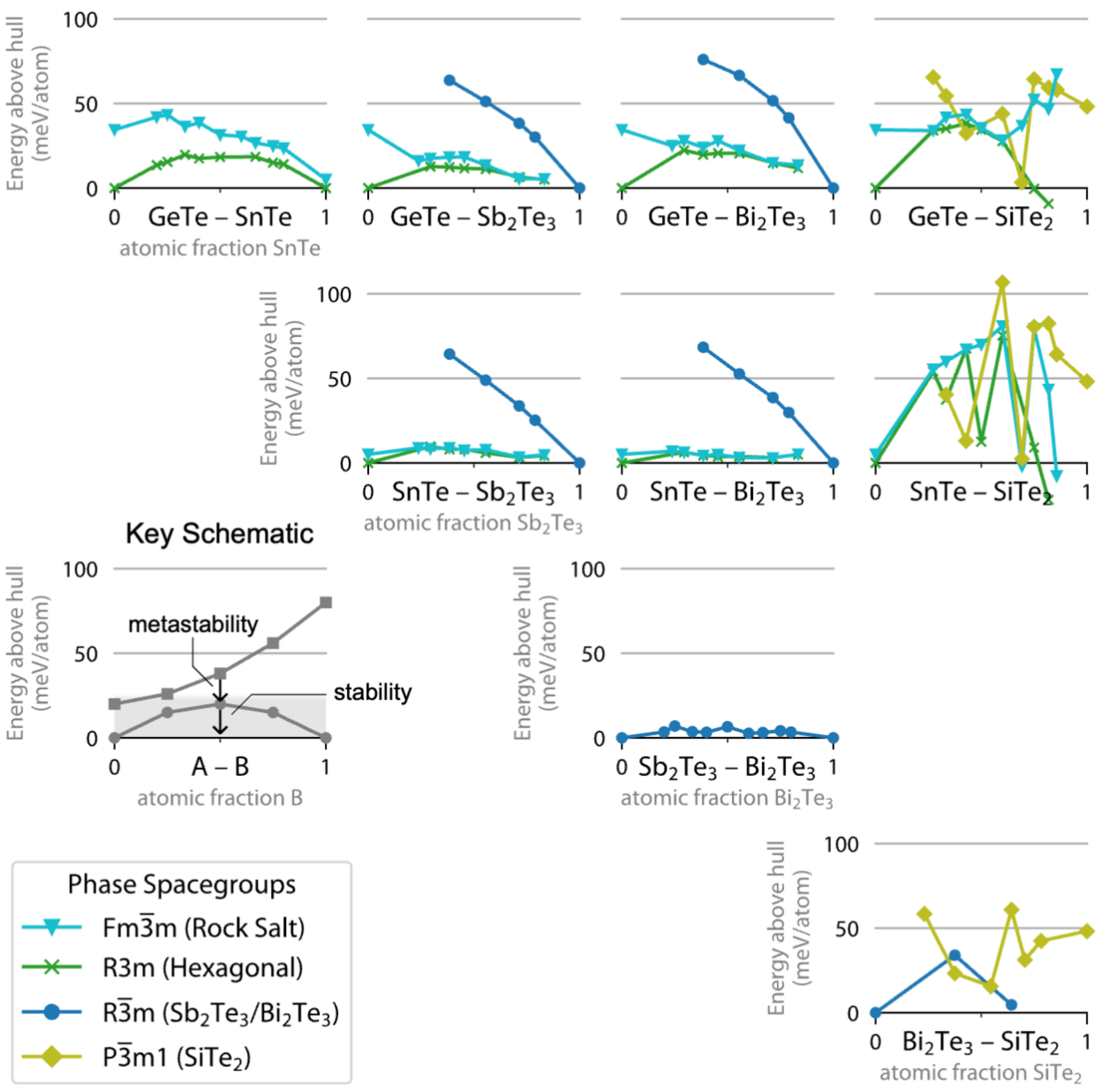

**Figure 3. The minimum DFT energies of ternary structures in each phase as a function of composition along telluride tie lines.** The schematic illustrates the stability and metastability of ternary phases with the 25 meV/atom cutoff (the gray shaded region) for stability, below which ternary phases are considered stable and miscible, and above which ternary compositions are expected to be unstable and phase segregate. Rows and columns correspond to the left and right binary compounds, respectively, ordered as: GeTe, SnTe, $Sb_2Te_3$, $Bi_2Te_3$, and $SiTe_2$. Horizontal axes are in units of atomic fraction of the right-hand composition.

The GeTe–$Bi_2Te_3$ mixture has the same polymorphs as GST. The hexagonal phase mixture energies are in the range of 15 to 25 meV/atom, RS phase mixtures are 15 to 35 meV/atom, and R$\bar{3}$m phase mixtures are much higher at 40 to 75 meV/atom. Similar to GST, our calculations indicate single-phase formation in this mixture, and a kinetically accessible RS phase 0 to 10 meV/atom higher in energy than the hexagonal phase. This agrees with experimental observations that this system has been synthesized without phase segregation, including compositions with $< 50\%$ $Bi_2Te_3$.[12] Experiments also find either the amorphous or RS phase as deposited and the hexagonal phase upon annealing[12]. Experiments also show this system has fast crystallization times similar to GST[12], consistent with our Ostwald-rule framework.

In the GeTe–SnTe mixture, the hexagonal phase is consistently below 25 meV/atom, while the energy of the ternary RS phase is about 25 to 50 meV/atom, decreasing monotonically as a function of SnTe fraction. Experimentally, GeTe and SnTe form a continuous solid solution, in agreement with the consistently low energy of the hexagonal phase[18]. As the SnTe fraction increases, the energy of the metastable RS polymorph decreases, indicating a kinetically more accessible crystallization pathway. This trend is consistent with crystallization measurements that adding SnTe significantly reduces the crystallization time of GeTe.[19]

In the $Sb_2Te_3$–$Bi_2Te_3$ mixture, the common polymorph, the R$\bar{3}$m phase, has a uniformly low energy $< 10$ meV/atom across the entire composition range, suggesting a solid solution. Indeed, a solid solution with the R$\bar{3}$m structure is observed experimentally [20]. No other polymorphs were reported in experiments with increasing quench rate [20], in agreement with the single low-energy phase predicted in our calculations (Fig. 3). Based on our framework, the absence of a low-energy metastable polymorph suggests that $Sb_2Te_3$–$Bi_2Te_3$ is unlikely to exhibit GST-like rapid crystallization. Consistent with this, optical PCM experiments with this mixture showed long laser irradiation times to achieve the crystalline state[21].

The last column of Figure 3 shows mixtures of GeTe, SnTe, and $Bi_2Te_3$ with $SiTe_2$. These candidate mixtures exhibit erratic DFT energies across the tie-line. $SiTe_2$ is predicted to decompose into Si and Te using the r$^2$SCAN functional, so its endpoint compound shows significant energy above the hull. The presence of low-energy compositions ($< -10$ meV/atom) among high-energy compositions ($> 50$ meV/atom) suggests that specific ordered structures are stable, but with limited compositional and structural flexibility. These materials are therefore unlikely to exhibit the phase behavior as anticipated from our framework.

For the SnTe–$Sb_2Te_3$ and SnTe–$Bi_2Te_3$ mixtures, both the hexagonal and RS phases have energies below 10 meV/atom, and their energy difference is negligible, suggesting that ternary compositions can form a single phase. In agreement, single-phase ternary compounds with the same layered structures as the GST compounds (Figure 1) are experimentally reported[22,23]. Nearby mixtures $Sn_xTe_{100-x}$ ($x < 0.35$) – $Sb_2Te_3$ have been reported with similar layered structures as GST and require less power to switch than GST[24]. Consistent with the Ostwald's-rule framework, these results suggest SnTe–$Sb_2Te_3$ and SnTe–$Bi_2Te_3$ are promising candidate systems that merit further experimental exploration as potential fast-switching PCMs.

## 2.4 Selenide Mixtures

Figure 4 shows the energies of different polymorphs in the candidate selenide mixtures. The selenides generally have more diverse structures and higher energies above hull than tellurides. Many parent binary structures have large primitive unit cells (e.g., 20 atoms for $Sb_2Se_3$), which limits the accessible compositions in our calculations.

The GeSe–$Sb_2Se_3$ mixture is the selenide analogue of GST. GeSe is predicted not to be on the energy convex hull by the r$^2$SCAN functional. All ternary polymorphs have energies above 40 meV/atom, indicating likely phase separation. GeSe–$Sb_2Se_3$ does not exhibit the favorable features of our framework for fast-switching PCM behavior, and has little appearance in the PCM literature. The trends of polymorph energies in GeSe–$Bi_2Se_3$ are similar to those of GeSe–$Sb_2Se_3$. The ternary compositions have energies above hull of 30 meV/atom or higher, indicating likely phase segregation.

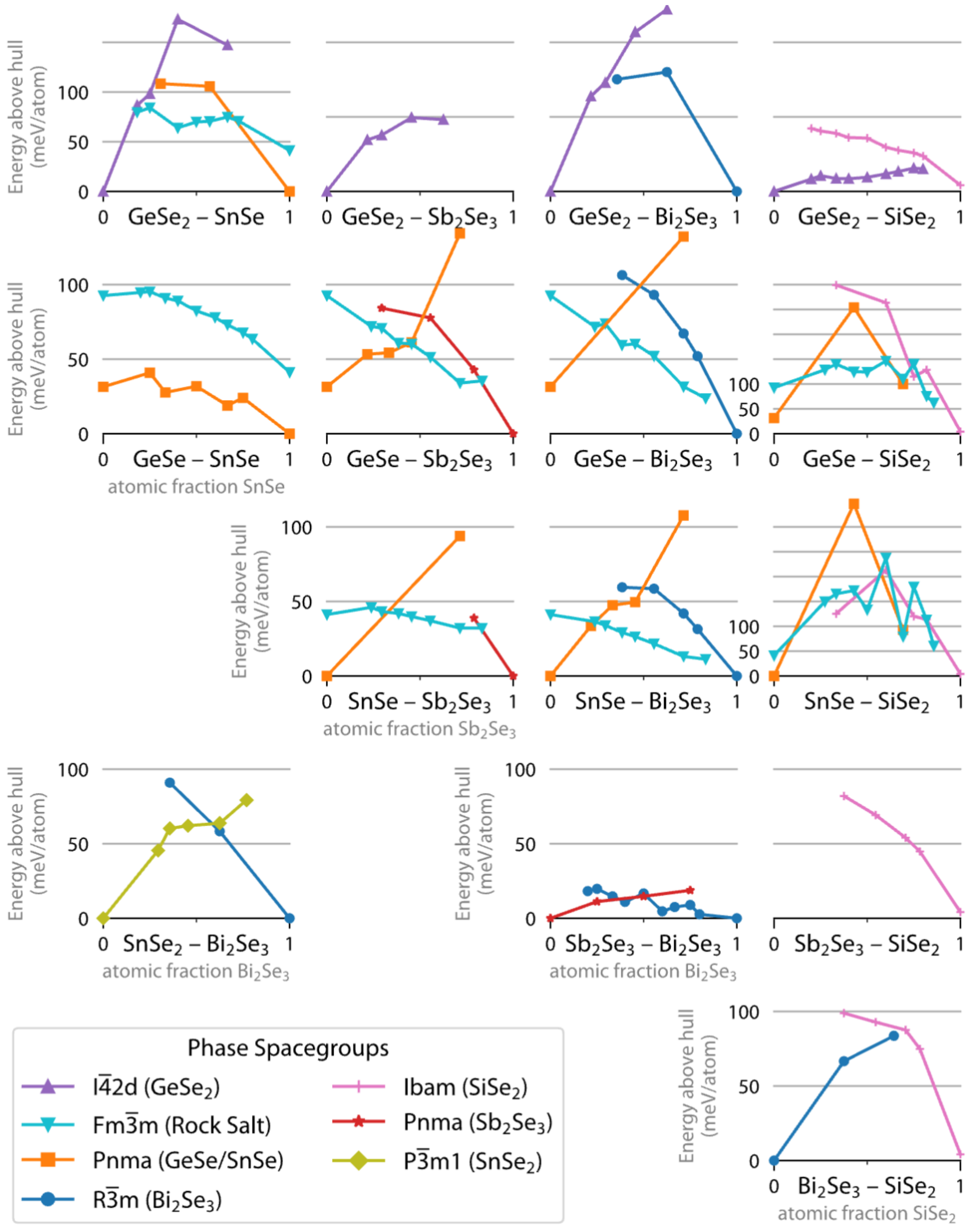

**Figure 4. The minimum DFT energies of ternary structures in each phase as a function of composition along selenide tie lines**. Rows and columns correspond to the left and right binary compounds, respectively, ordered as: $GeSe_2$, GeSe, SnSe, $Sb_2Se_3$, $Bi_2Se_3$, and $SiSe_2$. $SnSe_2$–$Bi_2Se_3$ is shown separately in the bottom left. Horizontal axes are in units of atomic fraction of the right-hand composition.

The SnSe–$Sb_2Se_3$ mixture also appears similar to the GeSe–$Sb_2Se_3$ mixture. SnSe shares the same orthorhombic and RS polymorphs as GeSe but is predicted to be on the convex hull by the $r^2$SCAN functional. The energies of some SnSe-based polymorphs are lower than in GeSe–$Sb_2Se_3$, though still above 40 meV/atom. Single-phase ternary compounds (e.g. $Sn_2Sb_2Se_5$) with orthorhombic structures have been found experimentally [25]. Overall, the SnSe–$Sb_2Se_3$ mixtures have higher polymorph energies than those sought by our thermodynamic framework.

For the GeSe–SnSe mixture, the ternary orthorhombic phase has energies of 25 to 50 meV/atom, while the RS phase has energies of 50 to 100 meV/atom. These results indicate that the orthorhombic phase may form a solid solution over part of the composition range, and the RS phase is likely too high in energy for fast crystallization. An experiment with $Ge_2Se_3$/SnSe stacks (similar to GeSe–SnSe mixtures) reported switching behavior when the two compounds were made to alloy, but the crystallization speed was not reported [26].

Among the diselenides, the $SnSe_2$–$Bi_2Se_3$ mixture is the only selenide candidate mixture with a hexagonal crystal family match, as in the tellurides, but its energy results are similar to those of other selenides. The ternary compositions have high energies, above 50 meV/atom, indicating phase separation. Indeed, experiments find that this mixture has limited solubility[27]. This system also does not show a low-energy polymorph sought by our framework.

The ternary compositions in the $GeSe_2$–SnSe, $GeSe_2$–$Sb_2Se_3$, and $GeSe_2$–$Bi_2Se_3$ mixtures all have high energies of 40 meV/atom or higher, indicating phase separation. Experimentally, $GeSe_2$ mixtures have mostly been studied for their glass-forming ability, and generally phase segregate upon crystallization[28–30], in agreement with the DFT results.

$SiSe_2$ mixtures show similar results to their $GeSe_2$ counterparts. SnSe–$SiSe_2$, $Sb_2Se_3$–$SiSe_2$, $Bi_2Se_3$–$SiSe_2$, and GeSe–$SiSe_2$ all have very high (> 50 meV/atom) energies for their ternary structures. These mixtures likely behave like most of the $GeSe_2$ mixtures, including a tendency toward phase segregation. This is consistent with an experimental study of Si doping in GST, in which Si was found to separate into a Si-rich phase.[31] Similarly, experimental studies of Si-doped $Sb_2Se_3$ find that Si phase-segregates upon recrystallization, forming a nanocomposite of $Sb_2Se_3$ grains in an amorphous Si-rich matrix,[32] consistent with the high predicted energies for similar $Sb_2Se_3$–$SiSe_2$ mixtures in our calculations. Although such Si-rich phase separation may reduce the reset power of PCM devices by acting as high-resistance microheaters[31] or enable tailoring of optical phase-change properties[32], this differs fundamentally from the GST-like behavior targeted by our framework for fast-switching PCMs.

The $GeSe_2$–$SiSe_2$ ternary compositions have much lower energies than the other mixtures with either $GeSe_2$ or $SiSe_2$. Like $Sb_2Se_3$–$Bi_2Se_3$, the two binaries have different ground-state polymorphs. The energy of the $GeSe_2$ polymorph (10 to 25 meV/atom) is consistently below that of the $SiSe_2$ polymorph (25 to 45 meV/atom). As suggested by these DFT results, a single-phase mixture based on the tetragonal $GeSe_2$ phase may form over a part of the tie line, with $SiSe_2$-derived polymorph close in energy over a composition range. Although this differs from GST in which a low-energy polymorph exists across the tie line, there is a composition range with competing low-energy polymorphs that may fit the picture suggested by our framework, and may merit further experimental investigation.

Unlike the $Sb_2Te_3$–$Bi_2Te_3$ mixture, which forms a single solid solution based on a common ground-state structure, the selenides $Sb_2Se_3$ and $Bi_2Se_3$ don't share the same ground-state phase. The ternary structures from both polymorphs of this mixture have energies consistently below 25 meV/atom, increasing as the composition deviates from their parent binaries. The calculated energies indicate that the favorable ground-state structure changes near the middle of the tie line, consistent with the experimental observation of the phase change near 70% $Bi_2Se_3$[33]. The DFT results suggest a change in the preferred ground-state structure across the tie line. The presence of two competing low-energy polymorphs is broadly consistent with the thermodynamic picture of our framework and may merit further experimental exploration.

Additional indium selenide mixtures were examined and are summarized in the SI. The InSe–$Bi_2Se_3$, InSe–$SiSe_2$, and $In_2Se_3$–$SnSe_2$ systems all exhibit high energies, indicating likely phase segregation. Notably, the $In_2Se_3$–$Bi_2Se_3$ system exhibits two low-energy polymorphs over a range of compositions, in agreement with our framework, suggesting a potential candidate for further experimental investigation.

Among these selenide mixtures, the $SnSe–Bi_2Se_3$ mixture possesses a low-energy RS polymorph with energies of approximately 20 meV/atom, comparable to GST. In experiments, single-phase ternary compounds (e.g., $Sn_4Bi_2Se_7$) have been reported with RS structures similar to those of GST [34]. Unlike GST, however, the calculations do not clearly indicate a low-energy phase below the RS structure as sought by our framework. Given that the RS phase is key to enabling rapid switching in GST, the RS-phase $SnSe–Bi_2Se_3$ mixture may merit further experimental investigation as potential PCMs.

Overall, compared to tellurides, the selenide mixtures have higher energies and are more likely to phase segregate. Selenides also show fewer polymorphs, and their polymorphs are generally higher in energy than tellurides. As a result, they generally do not exhibit the features targeted by our framework. These factors may help explain why ternary selenides are less common than ternary tellurides in the PCM literature.

## 2.5 Multi-Anion Mixtures

Finally, we present the results for mixtures of selenides and tellurides in Figure 5. We consider only mixtures in which the cation valence does not change, which excludes GeTe–$GeSe_2$ and SnTe–$SnSe_2$.

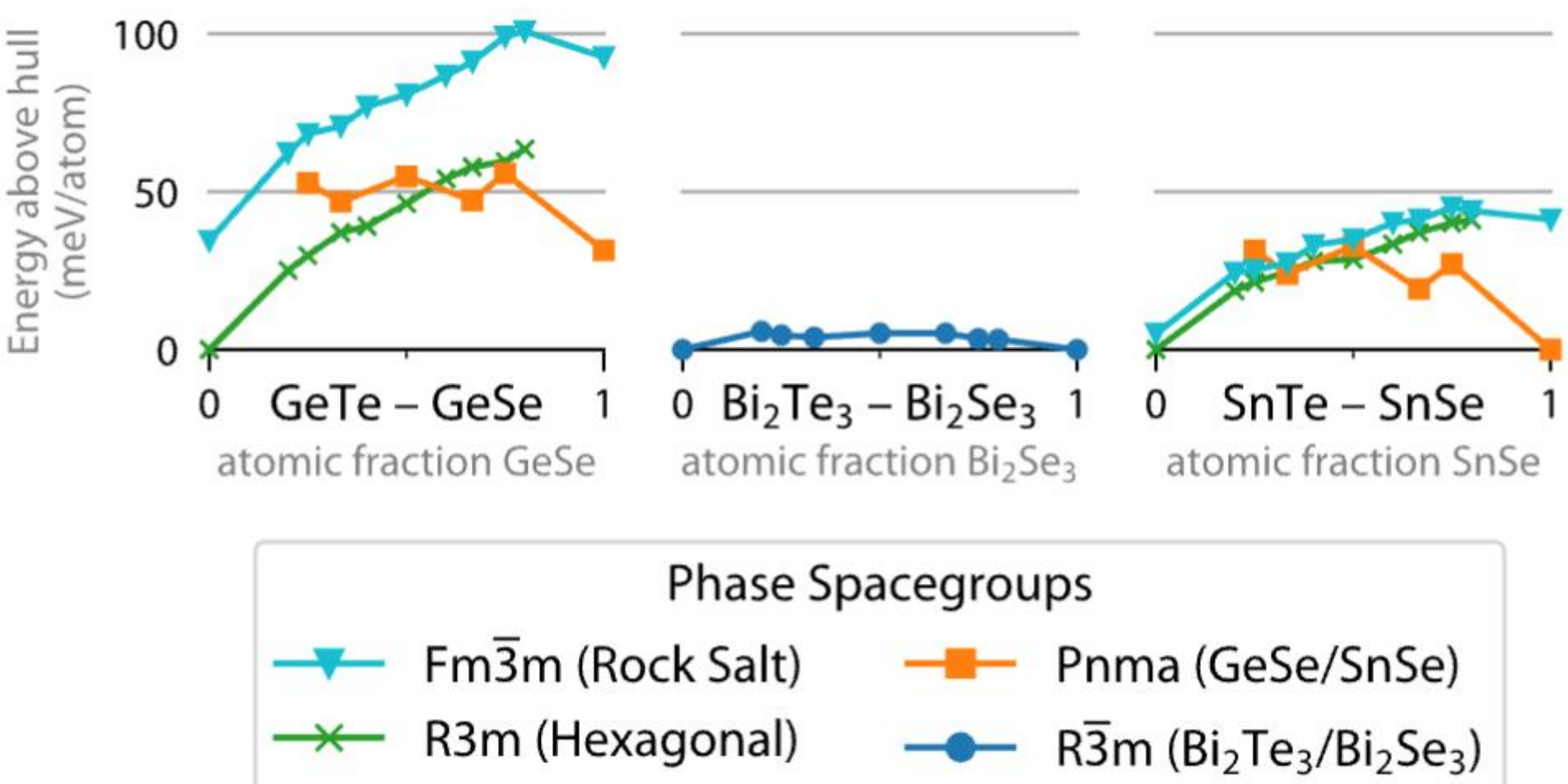

**Figure 5. The minimum DFT energies of ternary structures in each phase as a function of composition along multi-anion tie lines**. Horizontal axes are in units of atomic fraction of the right-hand composition.

For GeTe–GeSe, the energies of the ternary compositions are above 25 meV/atom due to the high energy of GeSe. The lowest-energy structure transitions from the hexagonal GeTe phase to the orthorhombic GeSe phase, while the RS phase is approximately 40 meV/atom above the hexagonal phase, reaching 60 to 100 meV/atom. This suggests that as the concentration of GeSe increases, the mixture is more likely to phase segregate and less likely to exhibit the RS phase. Experimentally, mixtures with up to 70% GeSe have been demonstrated, and the addition of GeSe significantly slows crystallization speed[19], in agreement with the increasing energy of the RS phase.

The $Bi_2Te_3$–$Bi_2Se_3$ energies are nearly identical to those of the $Sb_2Te_3$–$Bi_2Te_3$ system. These two binaries also have the same structure, and the ternary compositions between them all have low energies below 10 meV/atom, indicating a favorable solid solution. Like $Sb_2Te_3$–$Bi_2Te_3$, there is no metastable polymorph present to facilitate fast crystallization. $Bi_2Te_3$–$Bi_2Se_3$ solid solutions have been experimentally demonstrated for thermoelectric and optical applications[35], but don't appear in the PCM literature.

Like GeTe–GeSe, the SnTe–SnSe mixture shows a transition in the lowest energy phase between the hexagonal and orthorhombic phases, but the ternary compositions are much lower in energy (20 to 50 meV/atom). The energy of the RS phase is 25 to 50 meV/atom, only a few meV/atom above the hexagonal phase. These results suggest that this mixture may form the RS phase if the concentration of SnSe is not too high. This system exhibits the thermodynamic features targeted by our framework, although phase segregation may be likely due to the relatively high energies. In agreement, differential scanning calorimetry measurements of $Sn_xSe_{80-x}Te_{20}$ glasses showed multi-step crystallization, similar to GST[36], and phase segregation was also observed[36].

## 3 Discussion and Conclusion

In this work, we examined ternary chalcogenide mixtures from a thermodynamic perspective motivated by Ostwald's rule-based crystallization behavior in GST. Within this framework, fast-switching PCM candidates are expected to exhibit both a low-energy ground-state crystalline phase and a low-energy metastable polymorph, particularly a RS-like phase, that may provide a kinetically accessible intermediate during crystallization. The results for the GST system serve as validation of this framework. Our calculations correctly identify the hexagonal phase as the ground state and the RS phase as slightly metastable, in agreement with prior calculations and experiments. Similarly, the calculations reveal the favorable low-energy polymorph landscape of the GeTe–$Bi_2Te_3$ system, which is also consistent with fast-switching behavior observed experimentally. These thermodynamic analyses also correctly reproduce several known behaviors in mixed chalcogenide systems, including the ground-state structures, phase transitions, solid-solution formation, and tendency to phase segregate. The calculations also show that selenides generally lie higher in energy, are more prone to phase segregation, and less often exhibit favorable low-energy polymorphs, which may help explain why they are less common than tellurides in the PCM literature. These agreements suggest that first-principles thermodynamic analyses of the polymorph energetics provide a useful framework for rationalizing why some binary-binary tie-line mixture systems are more favorable than others and for identifying potential PCM systems.

Guided by this framework, several mixtures emerge as promising candidates for further exploration. In particular, SnTe–$Bi_2Te_3$ and SnSe–$Bi_2Se_3$ both exhibit low-energy RS phases comparable to those of known fast-switching systems, and SnTe–SnSe shows substantially lower RS energies than GeTe–GeSe. These systems are also experimentally accessible, as reported in the thermoelectric literature [23,34,37]. The $In_2Se_3$–$Bi_2Se_3$ system is also predicted to exhibit similar metastable polymorph behavior. These systems merit experimental evaluation for PCM-relevant properties, including crystallization kinetics, cyclability, and application-relevant optical or electrical contrast.

The trends observed across the composition space also suggest a broader structural design principle: binary parent phases with octahedral coordination environments similar to RS-derived structures are more likely to yield low-energy RS-like ternary polymorphs. This can be seen by comparing the SnSe–$Sb_2Se_3$ and SnSe–$Bi_2Se_3$ systems. In SnSe–$Sb_2Se_3$, both parent phases are orthorhombic and exhibit either different or strongly distorted octahedral coordination environments. In contrast, SnSe–$Bi_2Se_3$ contains the hexagonal $Bi_2Se_3$ parent phase, which has a more RS-like octahedral coordination environment. This structural similarity is consistent with the more favorable RS-like polymorph energies predicted for SnSe–$Bi_2Se_3$.

Future PCM discovery efforts may extend beyond the binary compositions studied here to mixtures whose parent compounds share compatible octahedral structural motifs, such as $TiTe_2$ with GeTe and $Sb_2Te_3$. The computational framework can also be extended to mixtures of three binaries or even higher component systems. For example, our results suggest that quaternary PCMs, such as GeTe–SnTe–$Sb_2Te_3$, combine mutually compatible binary mixtures and may exhibit favorable polymorph features.

While this work focuses on identifying single-phase materials, phase segregation can, in some cases, be used beneficially for tailoring PCM properties[32]. For example, recent work on Si-incorporated $Sb_2Se_3$ showed that phase segregation into nanocomposites can be exploited as a design strategy for tuning optical phase-change behavior. Although phase segregation is generally unfavorable within our GST-inspired framework for identifying fast-switching single-phase PCM systems, this requirement varies by the target application. This suggests a broader direction for future materials discovery, in which phase-segregating systems may also be explored intentionally when the goal is to engineer optical, thermal, or other properties rather than replicating GST-like fast switching mechanisms.

Several limitations of the current computational approach could be addressed in future work. First, the DFT calculations are performed at 0 K and do not explicitly include temperature effects. To incorporate temperature effects under experimental conditions, temperature-dependent descriptors[38] or statistical approaches based on Monte Carlo simulations may be conducted[39]. Second, the present analysis evaluates energies relative to the convex hull of crystalline phases, and the relative energies between amorphous and crystalline phases are not captured directly. Incorporating amorphous structures would provide additional insight, but obtaining reliable amorphous configurations generally requires computationally demanding molecular dynamics simulations. The present work focuses on thermodynamic polymorph landscapes and does not directly compute crystallization kinetics[40], although there is potential to extract some kinetic information from the energy distributions of the ordered ternary structures. Molecular dynamics simulations may also be used to directly study the phase transition mechanisms, including the nucleation and growth of the crystalline phase from the amorphous state[10], providing detailed mechanisms of crystallization kinetics in promising candidate systems identified by our framework. Application-relevant functional properties, including resistivity or optical contrast, are also not calculated. These remain important directions for future work.

# 4 Methods

## 4.1 Candidate Binary Selection

Candidate binary structures were obtained from the Materials Project (MP)[41]. We excluded non-stoichiometric compositions, such as $SbTe_2$, which approximates the Sb–Te eutectic composition $Sb_{30}Te_{70}$, because they are unlikely to behave similarly to the stoichiometric binary parent compounds GeTe and $Sb_2Te_3$ which form GST.

To identify whether a pair of binary compositions is connected via a tie line in the 0 K DFT-predicted phase diagram, we used a combination of graph search and tools from pymatgen[42]. We constructed the DFT phase diagram using the phase diagram tool from pymatgen[43,44] with all MP entries in the relevant composition space. We then converted the phase regions in the phase diagram into a graph, in which tie lines are edges connecting composition vertices. We next used the A* algorithm[45] to find the shortest path between the two binary composition nodes in the graph. If all compositions on the shortest path between the two binaries were collinear, the two binaries were considered to be connected by a tie line, and were retained as a candidate pair.

Some binary compounds, such as $SiTe_2$, GeSe, and $BiSe_2$, are only predicted to be on the convex hull using GGA functional calculations and are predicted to be unstable by $r^2$SCAN functional[46] calculations. We still considered these compositions.

### 4.2 Vacancy Charge Compensation

When interpolating between two binary compositions to produce ternary compositions, we automatically determined the appropriate vacancy defect reaction for charge balancing based on the oxidation states of the relevant species. If the dopant and substituted species have the same oxidation state, no charge compensation was required. If the oxidation state of the dopant species is greater in magnitude than that of the substituted species, then additional vacancies were added to the substituted species sublattice. Otherwise, vacancies were added to the other sublattice of the binary structure. We excluded systems with cations of mixed valence, because the resulting multiple possible substitutions would require a more complex ordering scheme. The resulting compositions of the ternary structures were limited to simple stoichiometric ratios with denominators less than 6 (e.g., $Ge_2Sb_2Te_5$ and not $Ge_7Sb_8Te_{19}$) and to compositions achievable with less than 100 atoms.

### 4.3 Ternary Structure Ordering

Given the ternary compositions and parent binary structures, we used EnumLib[47] via pymatgen to generate the ternary structures. The cation and anion sublattices of the parent structures were updated with the fractional occupation of the three species in the ternary composition. EnumLib was used to enumerate all the symmetrically distinct atomic orderings with the minimum supercell size. We determined the possible compositions and their required minimum supercell size using the procedure outlined in pseudocode in Figure 6. We considered only compositions with simple stoichiometric ratios with a denominator of less than 6. After ordering, we discarded composition-structure combinations with more than 1,000 ordered structures, which were computationally expensive for DFT calculations.

**Figure 6.** Algorithm pseudocode for determining ternary compositions and required supercell sizes, given a dopant species and a parent binary structure.

```
structures = {}
compositions = {}
sizes = {}

for all supercell sizes such that the number of atoms is < 100:
    for all defect reaction multiples that fit in the supercell:
        determine the number of each species
        calculate the corresponding dopant / substituted fraction
        if the fraction is simple and not in compositions:
            assign the fractional occupancy to the parent structure
            add the resulting unordered structure to structures
            add the composition to compositions
            add the supercell size to sizes
```

## 4.4 DFT Calculations

We performed the DFT calculations using a combination of Perdew–Burke–Ernzerhof (PBE)[48] and $r^2$SCAN[46] calculations. The ordered ternary structures generated by EnumLib were first relaxed using the PBE functional, after which single-point self-consistent $r^2$SCAN calculations were performed on the relaxed structures to obtain more accurate energies.

The DFT calculations were performed with the Vienna Ab initio Simulation Package (VASP)[49]. All calculations were performed using MP compatible VASP inputs generated by pymatgen[42]. The energy above hull values reported in this work were calculated with respect to the entries in the MP [41], and were also calculated using pymatgen[42].

## 5 Data Availability

All the data for this work is available at https://doi.org/10.5061/dryad.xd2547dxn

## 6 Code Availability

All the code for this work is available at

https://github.com/mogroupumd/pcm_tieline_hull

## 7 Acknowledgements

The authors acknowledge funding provided by the National Science Foundation (award DMR-2329087), supported in part by industry partners, as specified in the Future of Semiconductors (FuSe) program.

## 8 Author Contributions

**Felix Adams:** Conceptualization, Methodology, Software, Validation, Formal Analysis, Investigation, Data Curation, Writing – Original Draft, Visualization; **Ichiro Takeuchi:** Conceptualization, Supervision, Funding Acquisition; **Carlos A. Ríos Ocampo:** Conceptualization, Supervision, Funding Acquisition; **Yifei Mo:** Conceptualization, Methodology, Resources, Writing – Review & Editing, Supervision, Funding Acquisition.

## 9 Competing Interests

Authors declare no competing interests.